\documentstyle[preprint,prc,aps]{revtex}
\begin{document}
\draft

\title{Towards exotic nuclei
 via  binary reaction mechanism}
\author{N.V.Antonenko$^{1,2}$, A.K. Nasirov$^{1}$,
T.M. Shneidman$^{1}$, and V.D. Toneev$^{1}$}
\address{$^{1}$Bogoluibov Laboratory of
Theoretical Physics, Joint Institute for
Nuclear Research,\\ 141980 Dubna, Moscow Region, Russia\\
$^{2}$Institut f\"ur Theoretische Physik der
Justus--Liebig--Universit\"at,
D--35392 Giessen, Germany}
\date{\today}
\maketitle

\begin{abstract}
Assuming a binary reaction mechanism, the yield of isotopes near
the heaviest $N=Z$ neutron-deficit nucleus $^{100}$Sn is studied
with a microscopic transport model. The large influence of
nuclear shell structure and isotope composition of the colliding
nuclei on the production of exotic nuclei is demonstrated. It is
shown that the reaction $^{54}$Fe+$^{106}$Cd seems to be most
favourable for producing primary exotic Sn isotopes which may
survive if the excitation energy in the entrance reaction
channel is less than about 100 MeV.  In the case of large
differences in the charge (mass) numbers between entrance and
exit channels the light fragment yield is essentially fed from
the decay of excited primary heavier fragments.  The existence
of optimal energies for the production of some oxygen isotopes
in the binary mechanism is demonstrated for the
$^{32}$S+$^{197}$Au reaction.
\end{abstract}

\pacs{PACS number(s): 25.70.Gh, 25.70.Jj}

\section{Introduction}
Besides different applications, the production of both
neutron-deficit and neutron-rich nuclei became very important
for the nuclear structure problems and new experiments with
radioactive beams.  The study of exotic doubly-magic nuclei with
$N=Z$ and neighboring isotopes was of a long-term interest for
nuclear structure.  The lightest $N=Z$ nuclei are the most
stable due to the double closed shell but with increasing the
atomic mass they depart from the proton drip line and become
unstable against proton decay. The doubly-magic $^{100}$Sn
nucleus having the deficit of 18 neutrons is expected to be the
heaviest $N=Z$ nuclear system which is still bound. The
formation of the  $^{100}$Sn isotope and study of its properties
is quite important for the further shell model development and,
in particular, for the comparative study of proton-neutron
interactions occupying the same orbits.

In spite of considerable efforts, experimentalists failed for a
long time in attempts to produce $^{100}$Sn. Only recently two
groups \cite{GAN,GSI} have synthesized this intriguing isotope
using two different approaches. At GANIL \cite{GAN}, the
reaction $^{112}$Sn+$^{nat}$Ni at the intermediate energy 63
MeV/A was measured, while at GSI \cite{GSI}, the high-energy
(about 1.1 \ GeV/A) Xe beam was used. In the latter case, the Sn
nuclei were produced in high-energy fragmentation of the
$^{124}$Xe projectile and 7 events of $^{100}$Sn nuclei were
observed during 277 hours of beam time.  The fragmentation-like
reaction was employed  at GANIL as well, where 11 nuclei of
$^{100}$Sn were identified over period of 44 hours. Taking into
account the primary beam intensities, this difference in the
production rate shows that the cross section for $^{100}$Sn
production is higher by about two orders of magnitude at
intermediate energies. However, the reaction mechanisms at these
two energies are quite different. In fact, high-energy
fragmentation, which is considered traditionally in terms of a
participant-spectator model, does not work in a whole scale at
bombarding energies per nucleon of several tens of MeV. In
addition, there is compelling evidence (see for example
\cite{Sch94})  for  simple binary reaction dynamics dominated by
collective degrees of freedom. In other words, in this
transitional energy range, where reaction mechanism evolves from
dynamics of mean-field phenomena to a growing importance of
two-body nucleon-nucleon interactions, projectile and target
nuclei are able to largely preserve their identities and to form
a damped (but not fully) dinuclear system.  Nucleon exchange
which occurs in this dinuclear system will be influenced by
nuclear structure and, therefore, the yield of a particular
isotope may depend strongly on the combination of colliding
heavy ions.  The aim of this paper is to study how the yield of
isotopes near $^{100}$Sn is sensitive to the choice of colliding
nuclei and excitation energy of the dinuclear system formed.
Our consideration is based on assuming a binary character of
two-nuclei interaction and on its microscopic treating  within a
transport approach having a relevance to the intermediate energy
range of heavy-ion collisions.  The method of calculation will
be applied to the study of the production of different isotopes
of O as well. Therefore, we may analyse the pecularities of the
production of light and heavy exotic nuclei via the binary
reaction mechanism.

\section{Model}
\subsection{Model assumptions}
In the binary mechanism the interaction may be subdivided
roughly into three stages: First, at the impinging stage of
interaction, the colliding ions quickly lose some part of the
kinetic energy of their relative motion and form a dinuclear
system; then, this composite systems evolves in time exchanging
nucleons, energy and angular momentum and this interaction
terminates when the system decays into primary fragments. At the
final stage the excited fragments are de-excited by particle
emission resulting in the observed reaction products.  This
process is very similar to deep inelastic transfer reactions
taking place at lower energies (see for example
\cite{STW78,SCHH,F87,Vol}) but in our case a full dissipation of
the relative kinetic energy is not assumed.  Of course for
higher energies, the division of reaction into these stages can
be considered as the first approximation and the complete study
of whole reaction dynamics is preferable.  However, in order to
avoid cumbersome calculations and obtain some qualitative
results, in this paper we will find the total dissipated energy
and its partition between primary fragments using the
microscopic approach \cite{Adam1}. Then these data will be used
in the calculations of the charge (mass) distribution of
reaction products.

\subsection{Basic formalism and partition of excitation energy
between primary fragments}
The total Hamiltonian of a dinuclear system is
written as
\begin{equation}
\label{iniham}
\hat  H=\hat H_{rel}({\bf R},{\bf P})+
\hat H_{in}(\xi)+\delta \hat V({\bf R},\xi),
\end{equation}
where the Hamiltonian $\hat H_{rel}$ describing the relative
motion depends on
the relative distance ${\bf R}$ between
the centers of mass of the fragments
and the conjugate momentum ${\bf P}$.
In (\ref{iniham}) the quantity
$\xi$ is a set of relevant intrinsic variables.
The last two terms in (\ref{iniham}) describe the internal
motion of nuclei and the coupling between the collective and
internal motions.  The coupling term $\delta \hat V({\bf
R},\xi)$ leads to a dissipation of the kinetic energy into the
energy of internal nucleon motion.

Neglecting the residual nucleon-nucleon interaction,
whose effect will be included later in the equation for
the single-particle density matrix, we write the sum of the
last two terms in (\ref{iniham}) as a single-particle
Hamiltonian of a dinuclear system
\begin{eqnarray}
\hat {\cal H}({\bf R}(t),\xi)=
\sum_{i=1}^{A_0}\left( -\frac{\hbar ^2}{2m}\Delta _i+
 \hat  V_P({\bf r}_i-{\bf R}(t)) + \hat V_T({\bf r}_i)\right),
\end{eqnarray}
where $m$ is the nucleon mass, $A_0=A_P+A_T$ is the total number of
nucleons in the system, and $\hat V_P$ and $\hat V_T$ are
the single-particle potentials created by the projectile
and target-like nuclei.

In the second quantization representation, the Hamiltonian
$\hat {\cal H}({\bf R}(t),\xi)$ is written as
\begin{eqnarray}
\label{DNSsec}
\hat {\cal H}({\bf R}(t),\xi)&=&
\sum_{P} \tilde\varepsilon_P({\bf R}(t))
 a_P^+a_P +
\sum_{T} \tilde\varepsilon_T({\bf R}(t))
 a_T^+a_T\nonumber\\
&+&
\sum_ {P\ne P'} \Lambda^{(T)}_{PP'}({\bf R}(t))a_P^+a_{P'} +
\sum_{T\ne T'} \Lambda^{(P)}_{TT'}({\bf R}(t)) a_T^+a_{T'}
\nonumber \\
&+&\sum_{P,T} g_{PT}({\bf R}(t))(a_P^+
a_T + {\rm h.c.}).
\end{eqnarray}
Here, $P\equiv (n_P,j_P,l_P,m_P)$ and $T\equiv
(n_T,j_T,l_T,m_T)$ are the sets of quantum numbers
characterizing the single-particle states with the energies
$\varepsilon_{P(T)}$ in an isolated  projectile-like and the
target-like nuclei, respectively. The single-particle basis is
constructed from the asymptotic wave vectors of the
single-particle states of the noninteracting nuclei as shown in
Ref.~\cite{Adam1}.  For this basis set, the matrix elements in
(\ref{DNSsec}) are defined as
\begin{eqnarray}
\tilde {\varepsilon}_P ({\bf R}(t))&=& \varepsilon_P +
<P|V_T({\bf r})|P>, \nonumber\\
\tilde {\varepsilon}_T ({\bf R}(t)) &=& \varepsilon_T +
<T|V_P({\bf r}-{\bf R}(t))|T>, \nonumber\\
\Lambda^{(T)}_{PP'}({\bf R}(t)) &=& <P|V_T({\bf r})|P'>,
\nonumber\\
\Lambda^{(P)}_{TT'}({\bf R}(t)) &=&
<T|V_P({\bf r}-{\bf R}(t))|T'>,
\nonumber \\
g_{PT}({\bf R}(t))& =& {1\over2} <P|V_P ({\bf r}-
{\bf R}(t)) +V_T ({\bf r})|T>. \label{matel}
\end{eqnarray}
The nondiagonal matrix elements $\Lambda _{PP^{\prime }}^{(T)}$
($\Lambda _{TT^{\prime }}^{(P)})$ generate the particle-hole
transitions in the projectile (target) nucleus. The matrix
elements $g_{PT}$ are responsible for the nucleon exchange
between reaction partners.  Approximating the single-particle
wave functions by the wave functions in the finite square well
with the width depending on energy \cite{Adam2}, these matrix
elements are easily calculated.  Since the trajectory
calculation  shows that the relative distance between the
centers of the nuclei could not be less than the sum of their
radii, the tails of the single-particle potentials can be
considered as a perturbation disturbing the asymptotic
single-particle wave functions and their energies.

With the Hamiltonian (\ref{DNSsec}) whole reaction dynamics
can be studied and distributions in all observable
variables can be found. To avoid the cumbersome calculation of
primary distribution of the reaction products, we use
two step consideration.
At the first stage of the reaction
the average dissipated energy and its sharing between
the parts of dinuclear system
is calculated with the model \cite{Adam1} by solving
the equation for the single-particle density matrix $\tilde n$
\begin{eqnarray}
\label{eqdm}
i\hbar \frac {\partial \hat{\tilde n}(t)}{\partial t} =
[\hat {\cal H}({\bf R}(t)),\hat{\tilde n}(t)] -
 \frac {i\hbar}{\tau}[\hat {\tilde n}(t) -
 \hat {\tilde n}^{eq}({\bf R}(t))].
\end{eqnarray}
Here the residual interaction is taken into account in the
linearized form ($\tau$-approximation), and $\tilde n^{eq}({\bf
R}(t))$ is a local quasi-equilibrium distribution, i.e. a Fermi
distribution with the temperature $T(t)$.  With Eq.~(\ref{eqdm})
one may solve our problem in principle.  Following the procedure
of Ref.~\cite{Adam1} and neglecting the dependence of
single-particle potentials and levels on the charge (mass)
asymmetry, one can approximately find only the first and the
second moments of the mass (charge) distribution. The correct
calculation of whole mass (charge) distribution is difficult
computation problem with the formalism suggested in
\cite{Adam1}.  Therefore, at the second step of our
consideration the determined excitation energy and its partition
are used as inputs for the microscopic model \cite{Adam}, which
allows us in the simple manner to calculate the charge and mass
distributions formed due to the nucleon exchange between the
nuclei in the dinuclear system.  In the model \cite{Adam} the
average values of $R$ are taken as $1.16(A_P^{1/3}+A_T^{1/3})$
fm during the interaction time and only the motion in charge
(mass) asymmetry is considered, i.e., the dependence of
single-particle potentials and levels on the charge (mass)
asymmetry is taken into account.  Our analysis shows that this
two step consideration is reasonable to estimate the yield of
the exotic nuclei in the binary reaction mechanism.  We found
that the excitation energy $E^*$ is almost equally shared
between primary fragments in the reactions considered.  For the
collision energies higher than 15 MeV/A, the full dissipation of
the kinetic energy does not occur and the excitation energy in
the system increases more slowly with $E_{\rm c.m.}$. Thus,
experimentally \cite{Sch94} and theoretically \cite{Adam1} we
know that in the range 20-50 MeV/A roughly half of the relative
kinetic energy is transfered into the excitation energy.

\subsection{Charge-mass distribution of primary fragments}
In general, the production cross section for a primary fragment
$(Z,N)$ may be written as follows:
\begin{eqnarray}
\frac{d^2\sigma}{dZdN}=2\pi\lambdabar^2
\int\limits_{0}^{\infty}dt
\int\limits_{0}^{\infty}dJ\, J\,\, \Phi(J) \,\,
P_{ZN}(E^*_J,t) \,\, G(t) \ ,
\label{1}
\end{eqnarray}
where the function
$$\Phi(J)=\exp\left(-\frac{|\bar J-J|}{\Delta J}\right)$$
defines the angular momentum "window" with the width $\Delta J$
for the given reaction and
$$G(t)=\frac{1}{\tau_0}\exp(-t/\tau_0)$$
is the life-time distribution of the composite system
characterized by some specific interaction time
$\tau_0 \approx 10^{-21}{\rm s}$.
The quantity $ P_{ZN}(E^*_J,t)$ is the probability to find
the dinuclear system at the
moment $t$ in the state with mass $A=Z+N$ and charge $Z$ of one
of the primary fragments.  The excitation energy $E^*_J$ depends
on the bombarding energy $E_{kin}$, impact parameter (angular
momentum $J$) and dissipation rate. In the considered energy
range $E_{kin}=8-50$ MeV/A the angular momentum "window" $\Delta
J$ for forming  a particular primary isotope is rather narrow
10--20 $\hbar$ (especially for low excitation energy to be of
the most interest). So, in the first approximation
\begin{equation}
 \frac{d^2\sigma}{dZ \ dN} \approx 4\pi\lambdabar^2 \bar J
\Delta J\,  P_{ZN}(E^*_{\bar J},t_{int}),
\label{a2}
\end{equation}
where $t_{int}\approx 10^{-21}$s is the average interaction time
estimated for the collisions considered, $\bar J$ depends on
$E_{kin}$ and is smaller than the angular momentum in the
grazing collision. Thus, the relative yields of the isotopes are
defined by $P_{ZN}(E^*_{\bar J},t_{int})$. The value of $\bar J$
in (\ref{a2}) is defined from the trajectory calculation for
each kinetic energy by using the formalism presented in
Ref.~\cite{Adam1}. This gives the results which are close to the
classical trajectory calculations ~\cite{schmidt}.

In order to find $P_{ZN}(E^*_{\bar J},t_{int})$ characterizing
the temporal evolution of the multinucleon transfer process, a
dynamical model should be applied.  In this work we use a
 version of the transport approach developed in Ref.
\cite{Adam}. The main advantage of this version is that the
transfer process is treated  on a proper microscopic footing.
The consideration of proton and neutron transfers, which occur
simultaneously, is based on the use of single-particle level
scheme taking into account the isospin dependence of the
single-particle energies. This is in contrast with the other
dynamical approaches \cite{STW78,F87,R82} which studied the
nucleon exchange process with the smoothed (liquid drop like)
potential energy surfaces.

On the macroscopic level, the master equation for
$P_{ZN}(E^*_{\bar J},t)$ has the following
form \cite{Adam}:
\begin{eqnarray}
\frac{d}{dt}P_{ZN}(E^*_{\bar J},t)&=&\Delta^{(-,0)}_{Z+1,N} \
P_{Z+1\, N}(E^*_{\bar J},t)+
\Delta^{(+,0)}_{Z-1,N} \ P_{Z-1\, N}(E^*_{\bar J},t)\nonumber\\
&&+\Delta^{(0,-)}_{Z,N+1} \ P_{Z\, N+1}(E^*_{\bar J},t)+
\Delta^{(0,+)}_{Z,N-1}
\ P_{Z\, N-1}(E^*_{\bar J},t)
 \nonumber\\
&&-(\Delta^{(-,0)}_{Z,N}+\Delta^{(+,0)}_{Z,N}+
\Delta^{(0,+)}_{Z,N}+\Delta^{(0,-)}_{Z,N}) \
P_{Z\, N}(E^*_{\bar J},t),
\label{meq}
\end{eqnarray}
where transition probabilities  are defined as
\begin{eqnarray}
\Delta^{(\pm,0)}_{Z}=\frac{1}{\Delta t}
\sum\limits_{P,T}|g^Z_{PT}|^2 \
n^Z_{T,P}(\Theta_{T,P}) \ (1-n^Z_{P,T}(\Theta_{P,T})) \
 \frac{\sin^2(\Delta t(\tilde\varepsilon_{P_Z}-
 \tilde\varepsilon_{T_Z})/2\hbar)}{(\tilde\varepsilon_{P_Z}-
 \tilde\varepsilon_{T_Z})^2/4}
\label{cc}
\end{eqnarray}
and similar expression for $\Delta^{(0,\pm)}_{Z}$ with the
replacement $Z\to N$. The expressions (\ref{meq}) and (\ref{cc})
are elaborated by using the Hamiltonian (\ref{DNSsec}) (see
Ref.~\cite{Adam} for a detail). The transition probability
(\ref{cc}) follows the Eq.~(\ref{eqdm}).  Here, the matrix
elements $g_{PT}$ (\ref{matel}) are taken between
single-particle states of the projectile-like  ($P$) and
target-like ($T$) nuclei in the dinuclear system and include
both the nuclear and Coulomb (for protons) mean-field
potentials.  Since only the motion in charge (mass) asymmetry is
considered in this subsection, we take
$R=1.16(A_P^{1/3}+A_T^{1/3})$ fm to calculate $g_{PT}$ for each
$Z$ and $N$.  In (\ref{cc}) we use $\Delta t=10^{-22}$s$\,
<t_{int}$.  The single-particle occupation numbers $n_{P}(\Theta_P)$
and $n_{T}(\Theta_T)$ depend on the thermodynamical temperatures
$\Theta_{P}$ and $\Theta_{T}$ in  the projectile-like  and 
target-like nuclei, respectively.

Solving (\ref{meq}) with (\ref{cc}) at the initial conditions
$P_{ZN}(E^*,0)=\delta_{Z,Z_P} \  \delta_{N,N_P}$ and
$E^*=E^*(\bar J,E_{kin})$, the primary isotope distributions are
found for the certain interaction time $t_{int}$ keeping in mind
the relation (\ref{a2}).  Since the collisions with small
interaction times $\sim 10^{-21}$~s is mainly considered here,
the equal sharing of excitation energy between the parts of the
dinuclear system is used in accordance with the results of the
previous subsection.

\subsection{Secondary distribution}
If the primary isotope distribution is known, the secondary
(observable) isotope distributions can be estimated by applying
the statistical decay model to every excited primary fragment
\cite{Stat}. In the binary reactions one can expect the
competition of two processes determining the final yield of the
exotic isotopes. With increasing excitation energy the primary
yield of the exotic nuclei increases but the surviving
probability of these isotopes may be reduced in the subsequent
de-excitation process. However, the de-excitation process takes
two opposite roles. On the one hand, it reduces the
multiplicity  of the primary exotic isotopes obtained.  On the
other hand, it can increase the final yield of the exotic nuclei
due to the decay of heavier primary nuclei which are not exotic.
The last effect can be called as the feeding effect. Therefore,
for the production of the exotic nuclei, the choice of the
colliding nuclei and kinetic energy should supply the optimal
relationship between the primary isotope distribution and
de-excitation process.

\section{Results and discussion}
\subsection{Production of exotic light isotopes of Sn}
Let us start from the discussion of primary yield of the exotic
isotopes of Sn in different reactions.  In Figs.~1 and 2 the
calculated primary yield of isotopes produced in the
$^{58}$Ni+$^{106}$Cd reaction, which is considered as one of the
most promising combinations for $^{100}$Sn production, are
presented in the form of $Q_{gg}$-systematics ($\delta$ is
so-called "non-pairing" energy correction) \cite{Vol}.  The
calculation results are given for the equilibrium initial
temperatures $\Theta=3.5$ and $2.0$ MeV which correspond to the
excitation energy $200$ and $65$ MeV, respectively. It is seen
that the isotope yields follow $Q_{gg}$-systematics in both
cases, so the shell structure effects (inherent to the
transition probabilities (\ref{cc})) at least survives till the
temperature as high as $3.5$ MeV. Ranging over 6--8 orders of
magnitude, $P_{ZN}$ at the given temperature can be approximated
rather well by a single straight line (in the logarithmic
scale). However, the slope of this line does not coincide with
the initial temperature $T$ and even does not scale with $T$:
slopes in Figs.~1 and 2 differ by about $30 \%$.  One should
note that full thermodynamical equilibrium has not been assumed
in the microscopic transport model.  Therefore, the
$Q_{gg}$-systematics can not be considered as an evidence in
favour of full thermodynamical equilibrium in the evolution of a
dinuclear system.  Nevertheless, the $Q_{gg}$-systematics seems
to be a useful working method for estimating the yield of
isotopes far away from the line of stability.

The primary yield of exotic nuclei may be very sensitive to the
combination of colliding ions. As demonstrated in Fig.~3, the
use of $^{54}$Fe target instead of $^{56}$Fe in the reaction
with $^{106}$Cd  beam increases the primary yield of
neutron-deficit Sn isotopes by an order of magnitude. The yield
increases rapidly with the  excitation energy and this growth is
more pronounced for $^{100}$Sn than for heavier tin isotopes.
Note that subsequent de-excitation may suppress the primary
isotope yield at high $E^*$ (see below).

Similar trends are seen from the excitation functions for
various reactions presented in Fig.4. In addition, the results
for $^{64}$Zn+$^{106}$Cd and $^{40}$Ca+$^{106}$Cd collisions
turned out to be close numerically to those for
$^{58}$Ni+$^{112}$Sn and $^{58}$Ni+$^{106}$Cd (see also Figs.1
and 2) reactions.  Among all the cases considered, the
$^{54}$Fe+$^{106}$Cd reaction, which has the smallest neutron
number in the entrance channel, seems to give the largest
primary yield of $^{100}$Sn.

Reasons for a preference of the $^{54}$Fe+$^{106}$Cd
reaction are illustrated
in Fig.5, where the smoothed driving potential $U(Z)$ defined
by
\begin{eqnarray}
U(Z)=B_1+B_2+U_{12}(R_m)-B_0.
\end{eqnarray}
is shown.
Here $B_1$ and $B_2$ are the liquid drop binding energies of the
nuclei in the dinuclear system, $U_{12}(R_m)$ is the value of
the nucleus-nucleus potential at distance $R_m=R_1+R_2+0.5$ fm
\cite{Adan}.  The calculated results are reduced to the binding
energy $B_0$ of the compound nucleus.  Since the difference
between the moment of inertia of the dinuclear system including
$^{100}$Sn as one of the fragment and the moment of inertia of
the initial dinuclear system is very small in the reactions
considered, the driving potential can be taken for zero angular
momentum to analyse the energy thresholds for production of
$^{100}$Sn.  For reactions $^{54}$Fe+$^{106}$Cd and
$^{56}$Fe+$^{106}$Cd, $U(Z)$ is practically the same but the
energy barrier in the exit channel corresponding to $^{100}$Sn
is higher by $\sim 10$ MeV in the last case what dramatically
reduces the $^{100}$Sn yield. Comparing reactions on $^{54}$Fe
and $^{58}$Ni targets (lower part of Fig.5), one see that there
is no essential difference in the energy barriers for producing
$^{100}$Sn. However, the driving force $|dU(Z)/dZ|$ near the
entrance point is slightly larger for the case of $^{58}$Ni
target and this system moves towards smaller charge asymmetry
with higher probability.  Both arguments, high energy barrier
and large driving force, are valid for  $^{181}$Ta+$^{106}$Cd
collisions (upper part of Fig.5), so it is hardly ever possible
to produce $^{100}$Sn in this reaction.

Fig.~6 shows how the primary yield is modified by particle
emission. It was assumed that the excitation energy $E^*$ is
partitioned equally between primary fragments. Such partition
results from the analysis of experimental data for peripheral
collisions \cite{Kwiat} and is reproduced qualitatively in our
calculations \cite{Adam1}.  It is noteworthy that a such energy
sharing is in conflict with hypotheses of thermodynamical
equilibrium, but it is in agreement with particle-hole nature of
the interactions in the dinuclear system.  Experimental data
\cite{toke} shows that, for very short interaction times in the
deep inelastic transfer reactions, a large part of the
excitation energy belongs to the light fragments. With
increasing interaction time, the sharing of the excitation
energy is driven towards the thermal equilibrium limit but does
not reach it.  The isotope distribution of the de-excited
fragments is expected to be sensitive to the excitation energy
sharing.  As is seen in Fig.~6, the de-excitation process
changes the situation drastically: only neutron-deficit nuclei
with $E^* < 100$ MeV may survive. The practical interpretation
of this is that only very peripheral collisions or collisions in
which the main part of the excitation energy is in the light
fragment will be effective for producing these Sn isotopes.  As
shown in Fig.~7, the excited neutron-deficit Sn isotopes decay
mainly by charged particle emission. Thus, the direct use of the
Sn beam may be not the most effective way to reach the
$^{100}$Sn isotope. The last argument may turn out to be
important for production of this isotope in the high-energy
fragmentation method \cite{GSI}, as well.

\subsection{Production of oxygen isotopes}
The example considered above concerns the production of the
exotic nuclei in the reactions with the charge number of
projectile (target) close to the charge number of the exotic
nucleus under question.  In these reactions the de-excitation
process depletes the yield of the exotic nucleus.  Let us now
consider the production of oxygen isotopes in the reaction
$^{32}$S+$^{197}$Au. In this case the light nucleus in the
initial dinuclear system is far from O and the production of the
O isotopes follows a long path of nucleon transfers. As the
result, a large number of intermediate nuclei between S and O is
primarily produced.  The decays of these nuclei in the
de-excitation process can give rise to the O isotopes which are
of interest to us.  Therefore, in the $^{32}$S+$^{197}$Au
reaction one can expect a large feeding effect for the O
isotopes due to the de-excitation of the primary nuclei.

In Fig.~8 the primary yield of the $^{15}$O, $^{19}$O and
$^{21}$O isotopes is shown as a function of $E_{kin}$.  This
yield quickly grows with $E_{kin}$ up to the value of 20 MeV/A
and then some saturation occurs. In accordance with
$Q_{gg}$-systematics, the primary yield of $^{15}$O is smaller
then yields of $^{19}$O and $^{21}$O. The de-excitation process
strongly influences the excitations functions (Fig.~8).  The
calculated excitation functions are in reasonable agreement with
the experiment \cite{Pen}.  Due to emission of charged particles
and neutrons from nuclei heavier then the given O isotopes, its
secondary yield increases as compared to the primary one. This
feeding effect increases the yield of $^{15}$O by about two
orders of magnitude as compared to primary yield.  As is seen in
Fig.~8 the primary yields of $^{19}$O and $^{21}$O are close to
each other. The feeding effect increases the yield of $^{19}$O
much more then $^{21}$O.  In addition, the yield of the
neutron-rich isotopes as $^{21}$O is depleted due to the neutron
emission.  For heavy isotopes, for example $^{21}$O, the
secondary excitation function may have a maximum at certain
energy. Therefore, there is an optimal colliding energy for the
production of the given neutron-rich isotope.  The yields of
$^{15}$O and $^{19}$O isotopes reach the saturation at the
energy of about 20 MeV/A.  At lower values of $E_{kin}$ the O
yield is small due to the small primary yield and weak influence
of the feeding effect.  At higher energies the de-excitation
process starts to deplete the O yield because the centers of the
charge and mass distributions of the reaction products are
shifted to the lighter nuclei.  In this case the saturation
regime is reached  for the neutron-deficit isotopes, while the
yield of the neutron reach isotopes as $^{21}$O started to
decrease.  The optimal energy for producing light exotic nuclei
in the binary reactions seems to be between 15 and 25 MeV/A.
The most interesting results to be inferred from Fig.~8 are the
feeding effect for the reactions in which the exit channel is
far from the entrance channel and the existence of the optimal
colliding energies for the production of some isotopes.

\section{Summary}
In conclusion note that the binary mechanism is realized in the
contest of the microscopic transport model and that the yield of
exotic nuclei depends essentially on the shell structure of the
nuclear partners in the entrance and exit channels. Production
of the given isotope may be optimized by an appropriate choice
of colliding ions. In particular, for producing the
neutron-deficit $^{100}$Sn, the reaction $^{54}$Fe+$^{106}$Cd
seems to be the most effective in comparison with the reactions
$^{56}$Fe+$^{106}$Cd, $^{64}$Zn+$^{106}$Cd,
$^{58}$Ni+$^{106}$Cd, $^{58}$Ni+$^{112}$Sn and
$^{40}$Ca+$^{106}$Cd due to comparatively low energy barrier and
small charge drift towards $N/Z$-equilibrium. Only primary
neutron-deficit isotopes with the excitation energy less than
$\sim 100$ MeV may survive in a strong competition with charged
particle emission what puts additional limits to angular momenta
contributing to an observable yield of the given isotope.

For the reactions with a considerable difference in charge
(mass) numbers between entrance and exit channels, the feeding
effect is important in the yield of some exotic nuclei. Due to
the emission of a charged particle from nuclei heavier than the
isotope studied, the production of some exotic isotopes is
possible. For these reactions, there exists the optimal
colliding energy for the production of a specific isotope.

For more detailed predictions, the consideration of the all
stages (dissipation of the excitation energy, its sharing and
nucleon transfer) of interaction should be selfconsistently
included into the model.  This work is in progress now based on
\cite{STW78}. Further model development needs also some
experimental support. In addition to isotope, angular and
momentum distributions to control the dissipation process,
knowledge of excitation functions and energy dependence of the
isotope dispersion is of great interest for reactions under
discussion.

\acknowledgments
We wish to thank G.G.~Adamian, M.~Lewitowicz, Yu.~Penionzhkevich
and M.~Ploszajczak for useful discussions. The authors (N.A. and
V.T.) acknowledge the warm hospitality of the theory group of
GANIL, Caen where a part of this work was done within the
IN2P3-Dubna agreement N 95--28.  The author (N.A.) is grateful
to the Alexander von Humboldt-Stiftung for the support during
completion of this work.

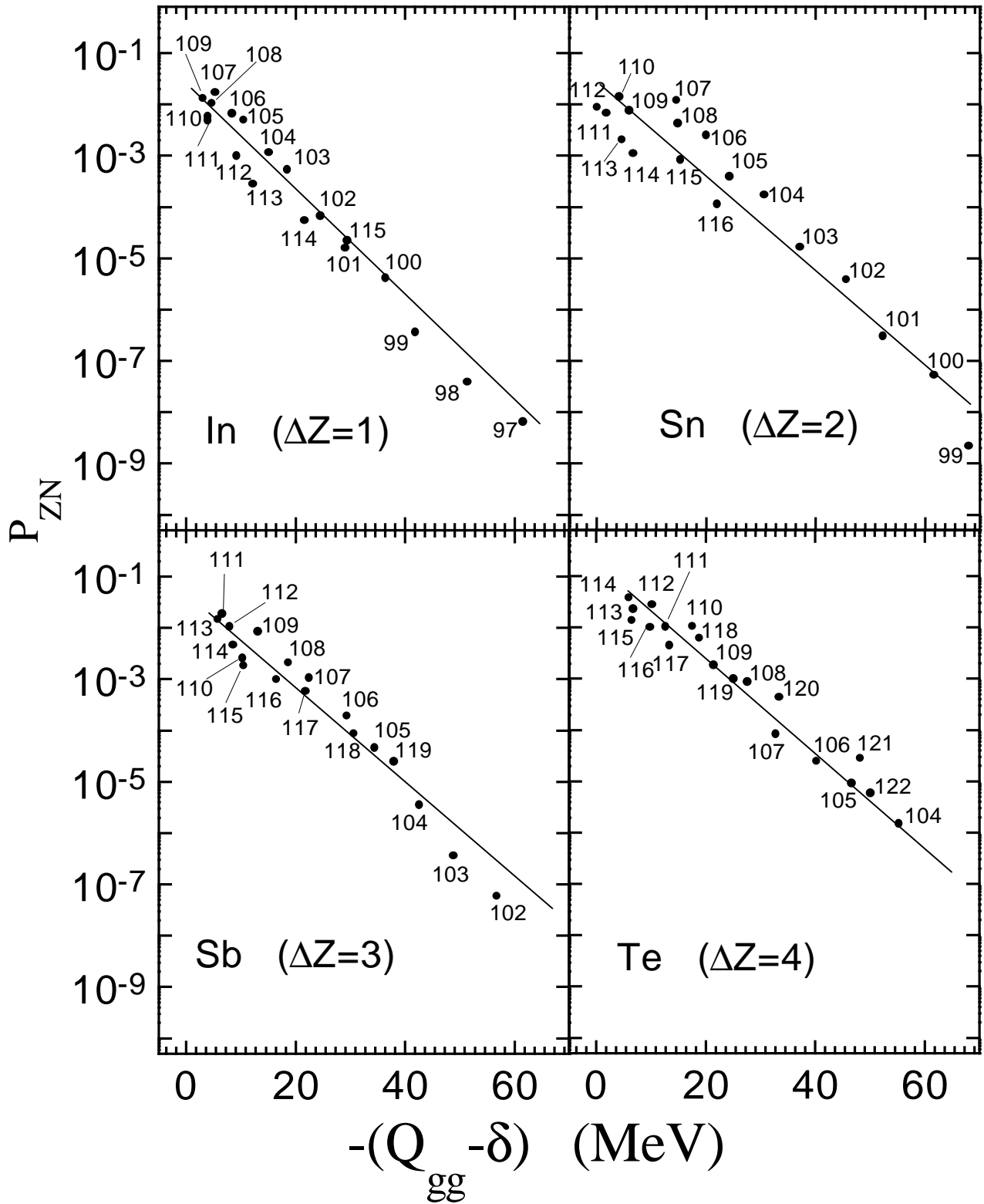
\begin{figure}
\caption{
The calculated $Q_{gg}$-systematics of primary yields for In,
Sn, Sb, and Te isotopes (points) from $^{58}$Ni+$^{106}$Cd
collision with the initial temperature $\Theta=3.5$ MeV. The solid
lines are drawn to guide the eye.}
\end{figure}

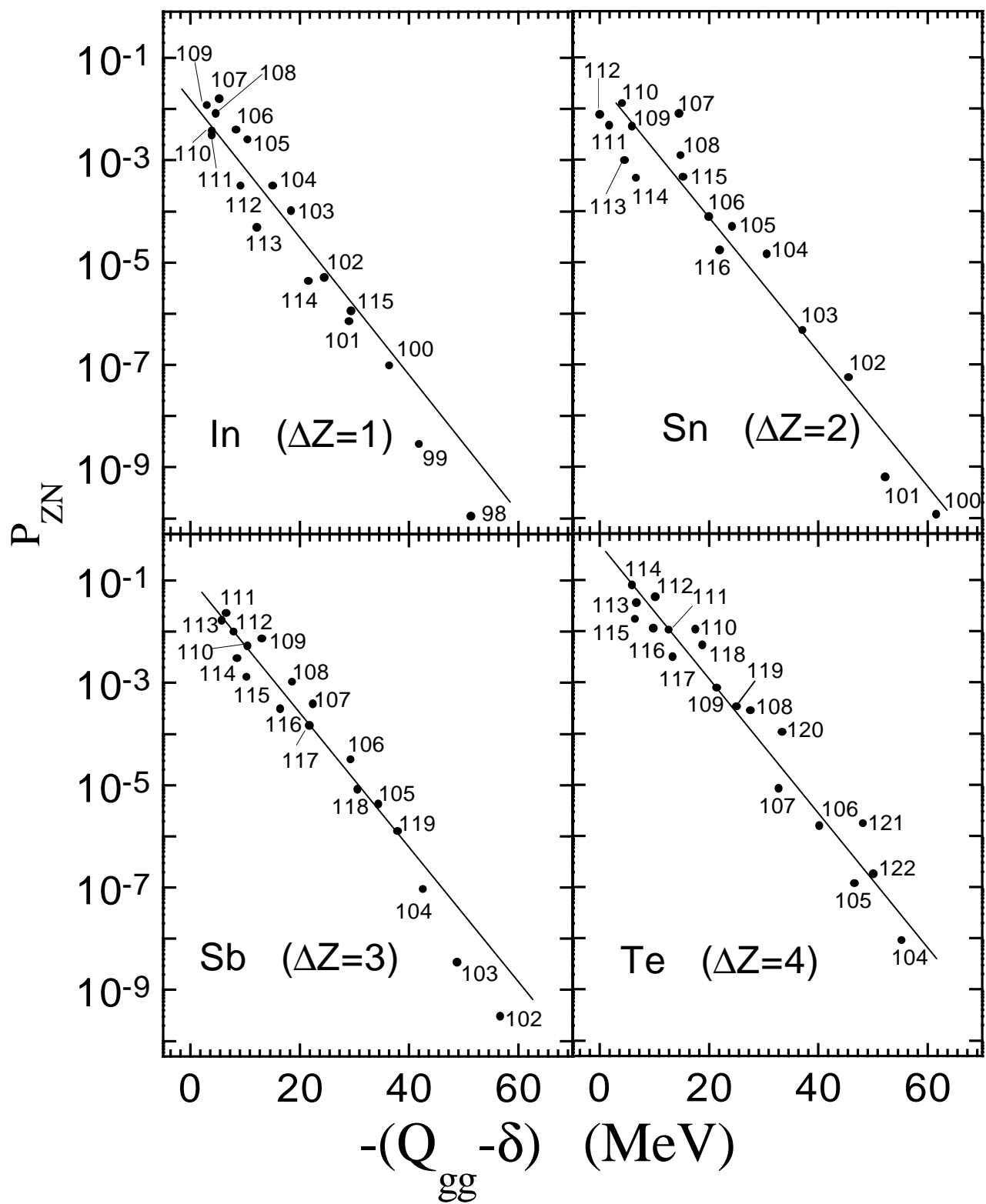
\begin{figure}
\caption{
The same as in Fig.1, but for $\Theta = 2.0$ MeV.}
\end{figure}

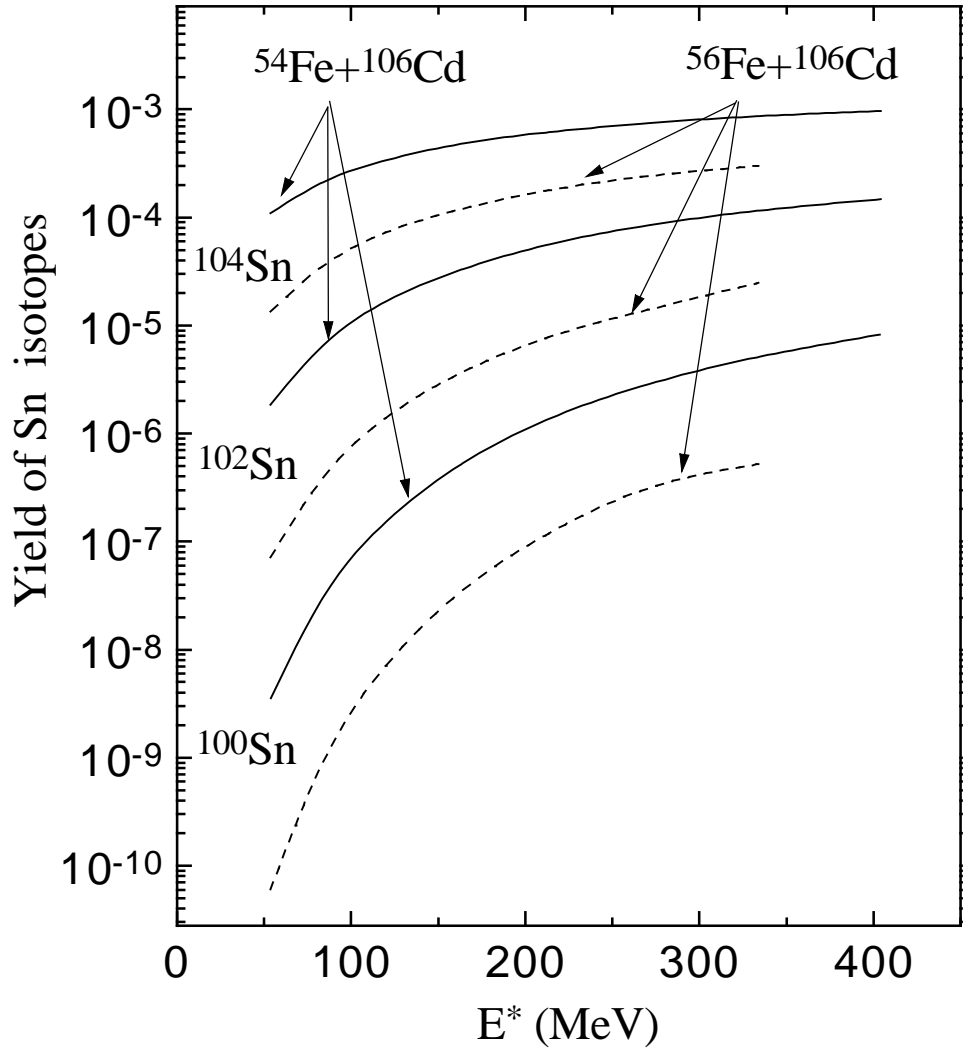
\begin{figure}
\caption{
The excitation functions of Sn isotopes from
$^{54}$Fe+$^{106}$Cd (solid lines) and  $^{56}$Fe+$^{106}$Cd
(dashed lines).}
\end{figure}

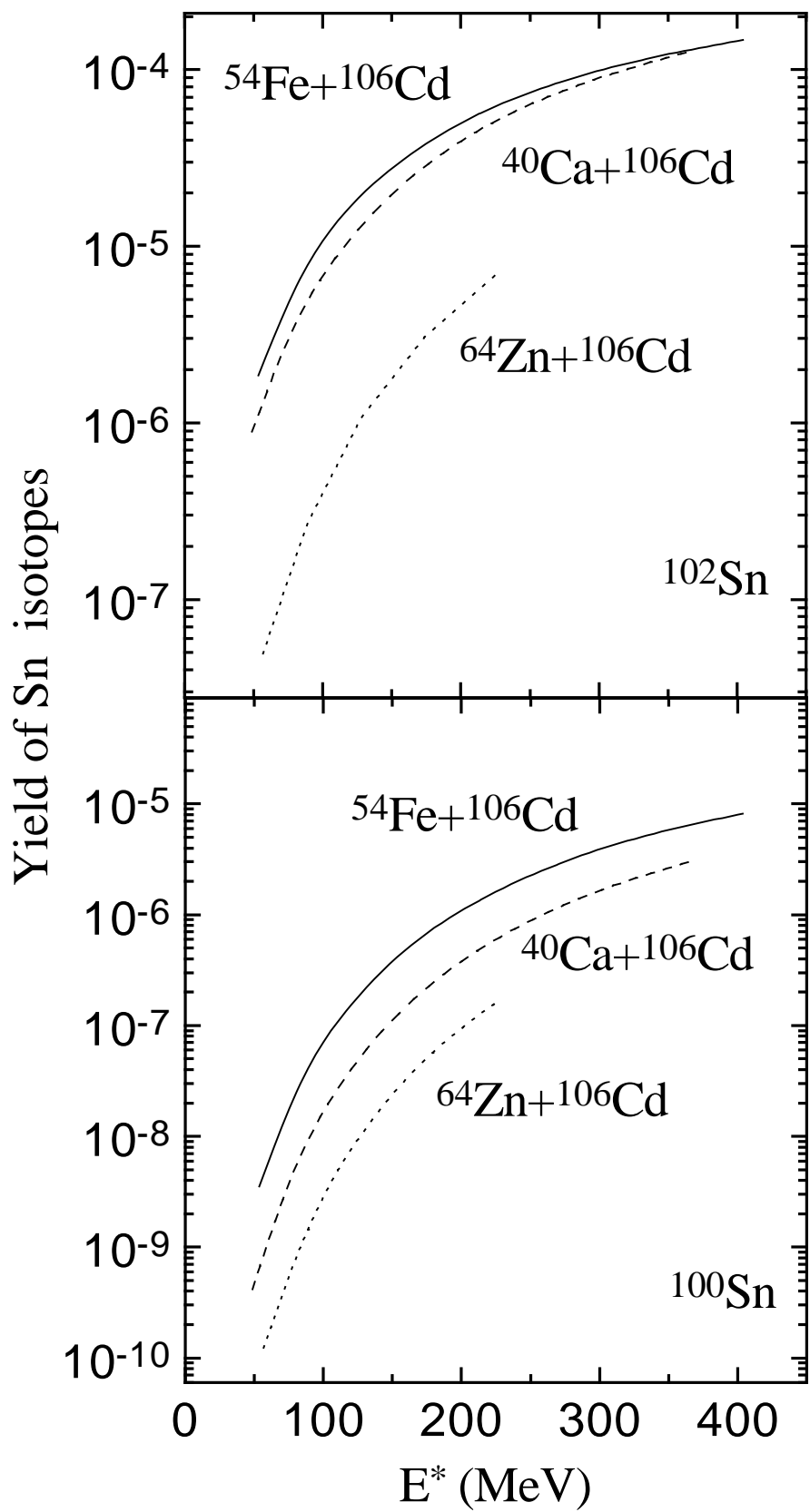
\begin{figure}
\caption{
The excitation energy dependence of $^{102}$Sn (upper part) and
$^{100}$Sn (lower part) for three composite systems:
$^{54}$Fe+$^{106}$Cd (solid lines), $^{40}$Ca+$^{106}$Cd (dashed
lines), and  $^{64}$Zn+$^{106}$Cd (dotted lines).}
\end{figure}

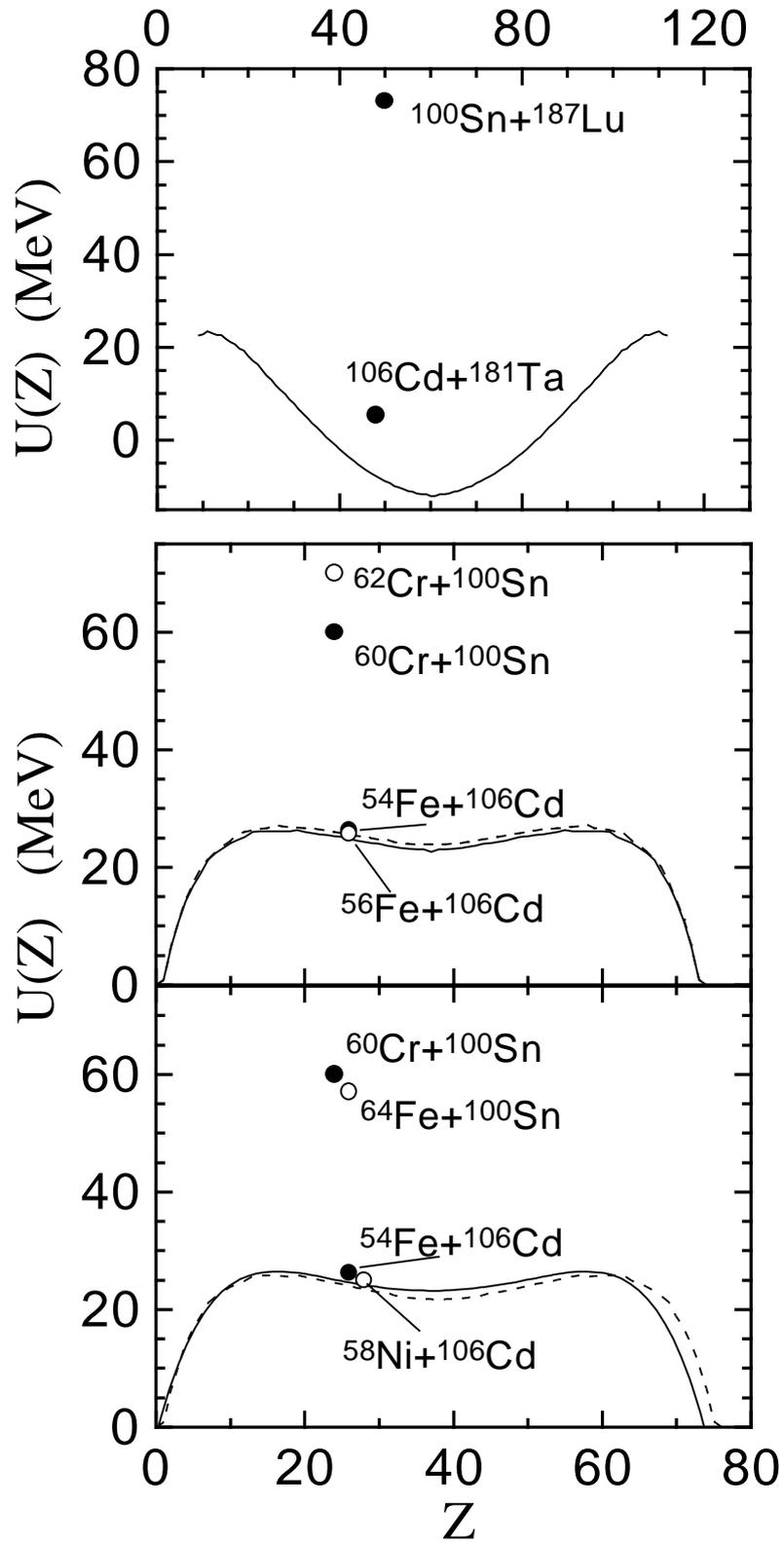
\begin{figure}
\caption{
The charge-asymmetry dependence of a smoothed part of potential
energy in $N/Z-$equilibrium for the following  systems:
$^{54}$Fe+$^{106}$Cd (solid lines), $^{56}$Fe+$^{106}$Cd (dashed
line, middle part),  $^{58}$Ni+$^{106}$Cd (dashed line, lower
part), and $^{181}$Ta+$^{106}$Cd (solid line, upper part).  The
points correspond to the system  energy in entrance channel and
to that in the final state with $^{100}$Sn isotope.}
\end{figure}

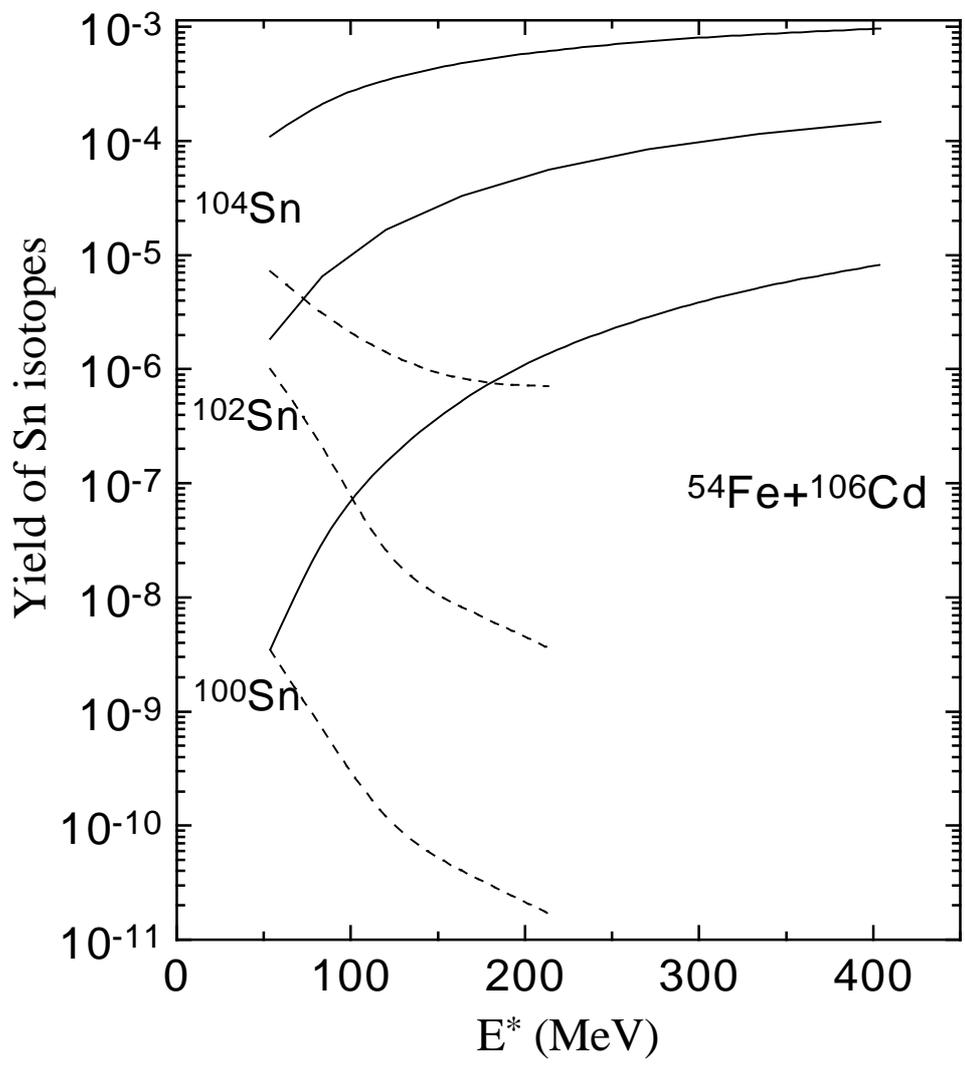
\begin{figure}
\caption{
The excitation functions of Sn isotopes from  the
$^{54}$Fe+$^{106}$Cd reaction.  The primary yield is shown by
solid lines, the dashed lines takes into account the depletion
due to de-excitation process.  The equal partition of excitation
energy is used in the calculations.}
\end{figure}

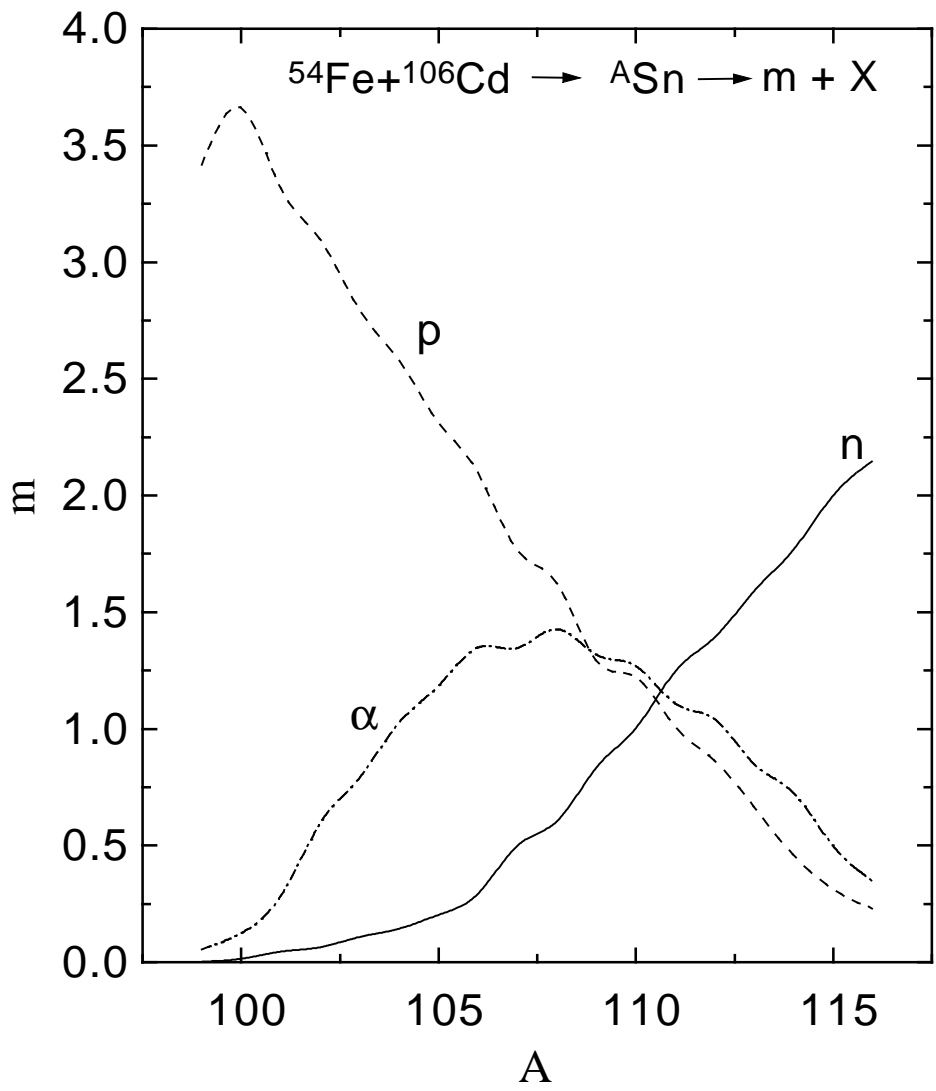
\begin{figure}
\caption{
Average multiplicity of particles (neutrons, protons and alpha
particles) emitted from the given primary Sn isotope produced in
the $^{54}$Fe+$^{106}$Cd reaction for initial excitation energy
$E^*=80$ MeV.}
\end{figure}

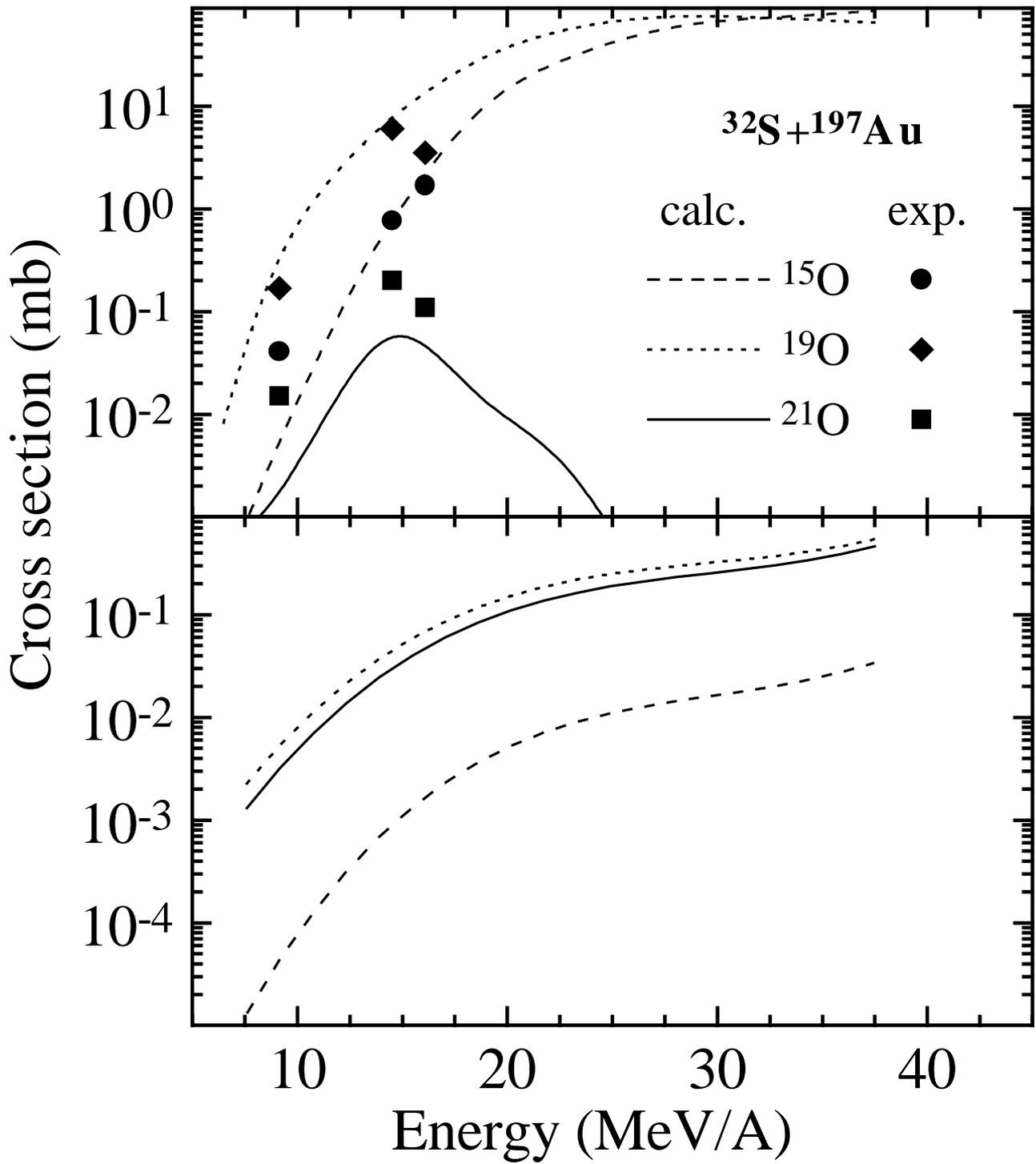
\begin{figure}
\caption{
 The primary (lower part) and
 secondary (upper part)
 yields of the $^{15}$O (dashed line), $^{19}$O (dotted line)
 and $^{21}$O (solid line) nuclei as a function of kinetic
 energy in
 the $^{32}$S+$^{197}$Au reaction.
 The experimental points are taken from  \protect\cite{Pen}. }
\end{figure}

\end{document}